\documentclass[fleqn,usenatbib]{mnras}
\usepackage{amsmath}
\usepackage{graphicx}
\usepackage{amssymb,txfonts}
\usepackage{xspace}
\usepackage{bm}
\usepackage[english]{babel}

\newcommand{\msun}{{\rm M}_{\sun}}
\newcommand{\rsun}{{\rm R}_{\sun}}

\newcommand{\source}{Cyg~X\babelhyphen{nobreak}3\xspace}

\title[Jets in the soft state in \source]{Jets in the soft state in \source caused by advection of the donor magnetic field and unification with low-mass X-ray binaries}

\author[X. Cao \& A. A. Zdziarski]
{Xinwu Cao$^{1,2}$\thanks{E-mail: xwcao@zju.edu.cn (XC); aaz@camk.edu.pl (AAZ)} and Andrzej A. Zdziarski$^{3}$\footnotemark[1]\\
$^1$Zhejiang Institute of Modern Physics, Department of Physics, Zhejiang University, 38 Zheda Road, Hangzhou 310027, China\\
$^2$Shanghai Astronomical Observatory, Chinese Academy of Sciences,
80 Nandan Road, Shanghai, 200030, China\\
$^3$Nicolaus Copernicus Astronomical Center, Polish Academy of Sciences, Bartycka 18, PL-00-716 Warszawa, Poland}

\date{Accepted 2019 December 04. Received 2019 November 21; in original form 2019 October 03}

\pagerange{\pageref{firstpage}--\pageref{lastpage}} \pubyear{2019}

\begin{document}

\maketitle \label{firstpage}

\begin{abstract}
The high-mass accreting binary Cyg X-3 is distinctly different from low-mass X-ray binaries (LMXBs) in having powerful radio and $\gamma$-ray emitting jets in its soft spectral state. However, the transition from the hard state to the soft one is first associated with quenching of the hard-state radio emission, as in LMXBs. The powerful soft-state jets in Cyg X-3 form, on average, $\sim$50\,d later. We interpret the initial jet quenching as due to the hard-state vertical magnetic field quickly diffusing away in the thin disc extending to the innermost stable circular orbit in the soft state, or, if that field is produced in situ, also cessation of its generation. The subsequent formation of the powerful jets occurs due to advection of the magnetic field from the donor. We find this happens only above certain threshold accretion rate associated with appearance of magnetically driven outflows. The $\sim$50\,d lag is of the order of the viscous timescale in the outer disc, while the field advection is much faster. This process does not happen in LMXBs due to the magnetic fluxes available from their donors being lower than that for the wind accretion from the Wolf-Rayet donor of Cyg X-3. In our model, the vertical magnetic field in the hard state, required to form the jets both in Cyg X-3 and LMXBs, is formed in situ rather than advected from the donor. Our results provide a unified scenario of the soft and hard states in both Cyg X-3 and LMXBs.
\end{abstract}

\section{Introduction}\label{intro}

There are several spectral states in black-hole (BH) X-ray binaries (XRBs) \citep*[e.g.,][]{1995ApJ...442L..13M,1997ApJ...477L..95Z,zg04, 2005A&A...440..207B, 2009ApJ...701.1940Y, 2010LNP...794...53B}, which probably correspond to different accretion modes \citep*{1997ApJ...489..865E}. In low-mass BH XRBs (LMXBs), relativistic jets are present only in certain states. Specifically, steady compact jets are associated with the low/hard spectral state, episodic jets are observed in the intermediate state, and jets are strongly suppressed in the soft state \citep*{1999ApJ...519L.165F, corbel01, 2003MNRAS.343L..99F, 2003MNRAS.344...60G, fender04}.

Relativistic jets are powered by taping energy from spinning BHs or/and the accretion disc with co-rotating magnetic field \citep{1977MNRAS.179..433B, 1982MNRAS.199..883B}. Numerical simulations have shown that a net poloidal/vertical magnetic flux is a necessary prerequisite for sustaining strongly magnetized accretion discs \citep{2016MNRAS.460.3488S}. One of promising mechanisms for the origin of a large-scale magnetic field  accelerating jets/outflows in X-ray binaries is advection of an external weak field by the discs \citep{1974Ap&SS..28...45B,1976Ap&SS..42..401B}. However, magnetic diffusion inevitably reduces the field advection in a turbulent accretion disc. The magnetic Prandtl number $P_{\rm m}=\nu/\eta$, where $\nu$ is the viscosity and $\eta$ is magnetic diffusivity, has been estimated to be $\sim$1 \citep{1979cmft.book.....P}. This agrees with numerical simulations, which found $P_{\rm m}\approx 1$ \citep*{2003A&A...411..321Y, 2009A&A...507...19F,2009ApJ...697.1901G}, or $P_{\rm m}\approx 2$--5 \citep{2009A&A...504..309L}. Thus, the magnetic flux diffuses away from the disc on the viscous time scale. Therefore, the field advection has been found to be rather inefficient in conventional turbulence-driven geometrically thin discs \citep*{1994MNRAS.267..235L}. This result seems consistent with the observational feature of jets switched off in the soft state. 

However, there is an exceptional case of \source. This high-mass XRB is the only known binary in the Galaxy that consists of a Wolf-Rayet (WR) star \citep{1992Natur.355..703V, 1996A&A...314..521V, 1999MNRAS.308..473F} and a compact object. The compact object is probably a BH, see, e.g., \citet*{zmb13}, \citet{koljonen17}, which presence we will hereafter assume. Both resolved jets with different size scales up to the projected distance of $\sim$1 light day \citep{2001ApJ...553..766M, 2004ApJ...600..368M, 2017MNRAS.471.2703E} and strong radio flares \citep*[e.g.,][]{szm08,koljonen10,koljonen18} have been observed in its {\it soft\/} spectral state. This shows that relativistic jets can also present in the soft state, which implies that an external field can be dragged inwards efficiently in a thin accretion disc at least in this source. Furthermore, cross-correlation between 15 GHz radio and soft X-rays in \source has shown a distinct radio lag of $\sim$50 d \citep{z18}, which is also a unique feature among XRBs. 

There are several possible ways to make efficient field advection in a thin disc. The large-scale magnetic fields can be dragged inwards in form of discrete, asymmetric patches in the disc, which leads to more efficient inward drift of the field \citep{2005ApJ...629..960S}. It was suggested that the field can be advected inwards efficiently by a hot tenuous gas above the disc \citep*{2009ApJ...707..428B, 2009ApJ...701..885L, 2012MNRAS.424.2097G, 2013MNRAS.430..822G}, because of the relatively larger radial velocity of the hot corona compared to the velocity at the midplane of the disc. However, \citet{2018MNRAS.473.4268C} found that maximal power of the jets accelerated by the field advected by the hot corona is always low, $\la$0.05 of the Eddington luminosity even for an extreme Kerr BH. 

An alternative mechanism was proposed by \citet{2013ApJ...765..149C}, in which magnetically driven outflows remove most of the angular momentum of the disc. The radial velocity of the disc is therefore substantially increased, and a weak external field can be advected inwards to form a strong field in the inner disc, which may then accelerate relativistic jets near the BH. Such mechanism was employed to explain jet formation in radio-loud quasars \citep{2016ApJ...833...30C}. In the case of XRBs, the detailed calculations show that the magnetic field strength at the outer edge of the disc is required to be as high as a few hundred G \citep{2019MNRAS.485.1916C}, which may be higher than the field advected from the companion stars in most XRBs. In LXMBs, the donors fill their Roche lobes and accretion proceeds via the inner Lagrangian region. Thus the magnetic flux can advected from only a small part of the stellar surface (cf.\ \citealt{sadowski16}). Indeed, the jets in BH LMXBs are seen to be switched off in the soft state. In the case of \source, it is known to contain a WR donor. Although the field strength at the surface of this WR star has not been measured, spectropolarimetric observations show that the magnetic fields of some WR stars can be as high as several hundred G \citep{2014ApJ...781...73D,2016MNRAS.458.3381H}. Furthermore, accretion is via the focused stellar wind, which can tap the magnetic flux from a large fraction of the stellar surface. Then, the magnetic field advected to the outer edge of the disc can be sufficiently strong to drive outflows from the thin disc, which significantly increase its radial velocity. As a result, the field can be sufficiently enhanced in an inner region of the disc to be able to accelerate jets near the BH.

In this work, we use the measured time delay of the radio emission with respect to the soft X-ray emission to constrain the field advection in the accretion disc in \source. We first summarize the observational features in Section \ref{observation}. In Section \ref{model}, we describe our model calculations and provide a unified scenario for both \source and BH LMXBs. We discuss our results in Section \ref{discuss}, and summarize the main conclusions in Section \ref{conclusions}.

\section{Observational features of Cyg X-3}\label{observation}

\begin{figure}
\centerline{\includegraphics[width=7.cm]{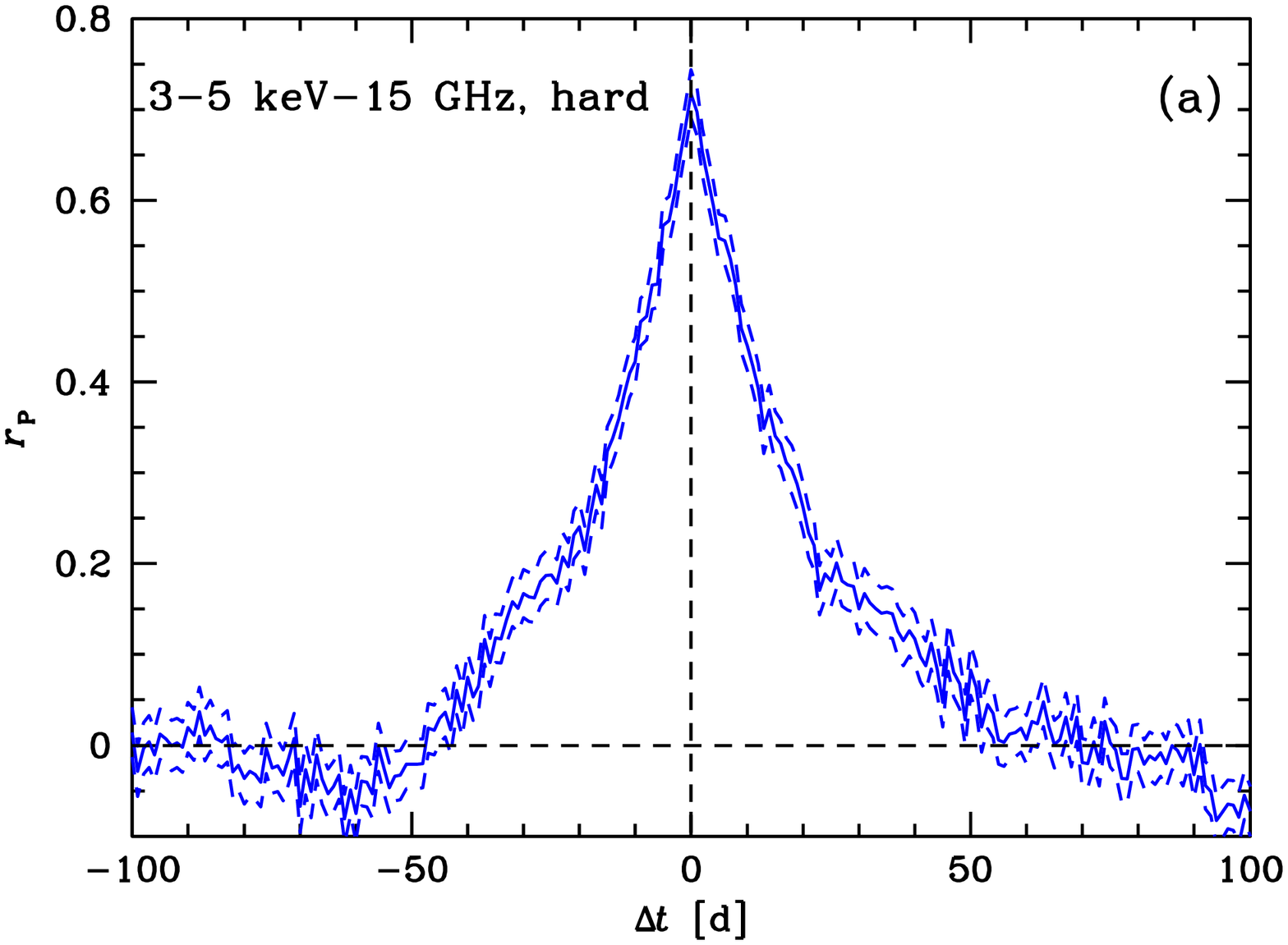}}
\centerline{\includegraphics[width=7.cm]{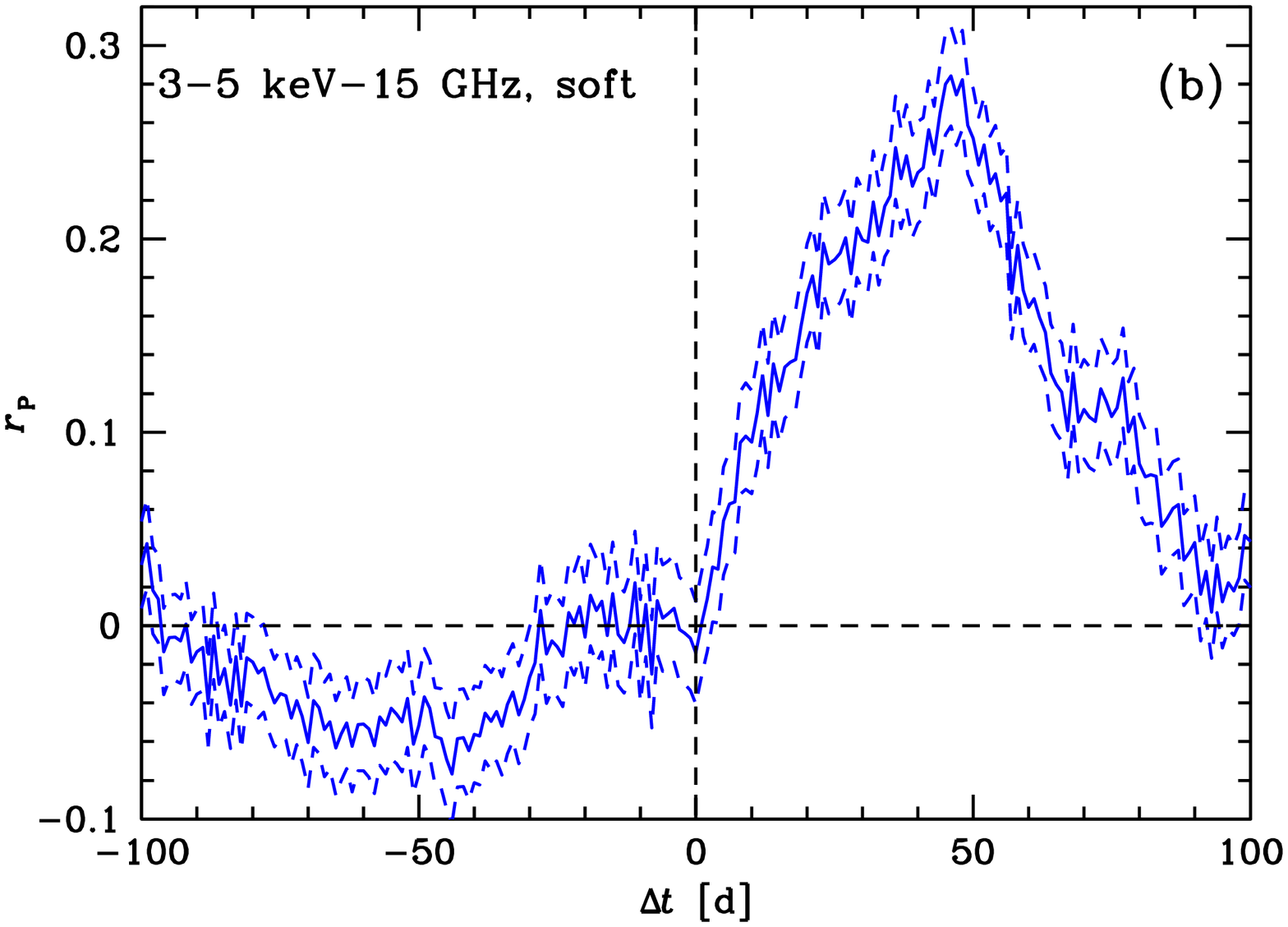}}
\caption{The cross-correlations (middle solid lines) between the 3--5 keV soft X-ray and the 15 GHz radio fluxes in the (a) hard and (b) soft and intermediate spectral states of \source. The dashed lines show the $1\sigma$ uncertainty ranges. The X-ray data are from the All-Sky Monitor \citep*{brs93} on board {\it Rossi X-ray Timing Explorer}. The radio data are from the Ryle and AMI telescopes \citep{pf97}. The hard (soft/intermediate) state is assumed to correspond to the 3--5 keV count rate of $<$ ($>$) 2.7 s$^{-1}$. In addition, the hard state is defined to correspond to the range of the radio flux of 30--300 mJy, while no constraint on it is imposed in the soft/intermediate state. Adapted from \citet{z18}, see that paper for details of the data and the used method. 
\label{corr} }
\end{figure}

\begin{figure}
\centerline{\includegraphics[width=7cm]{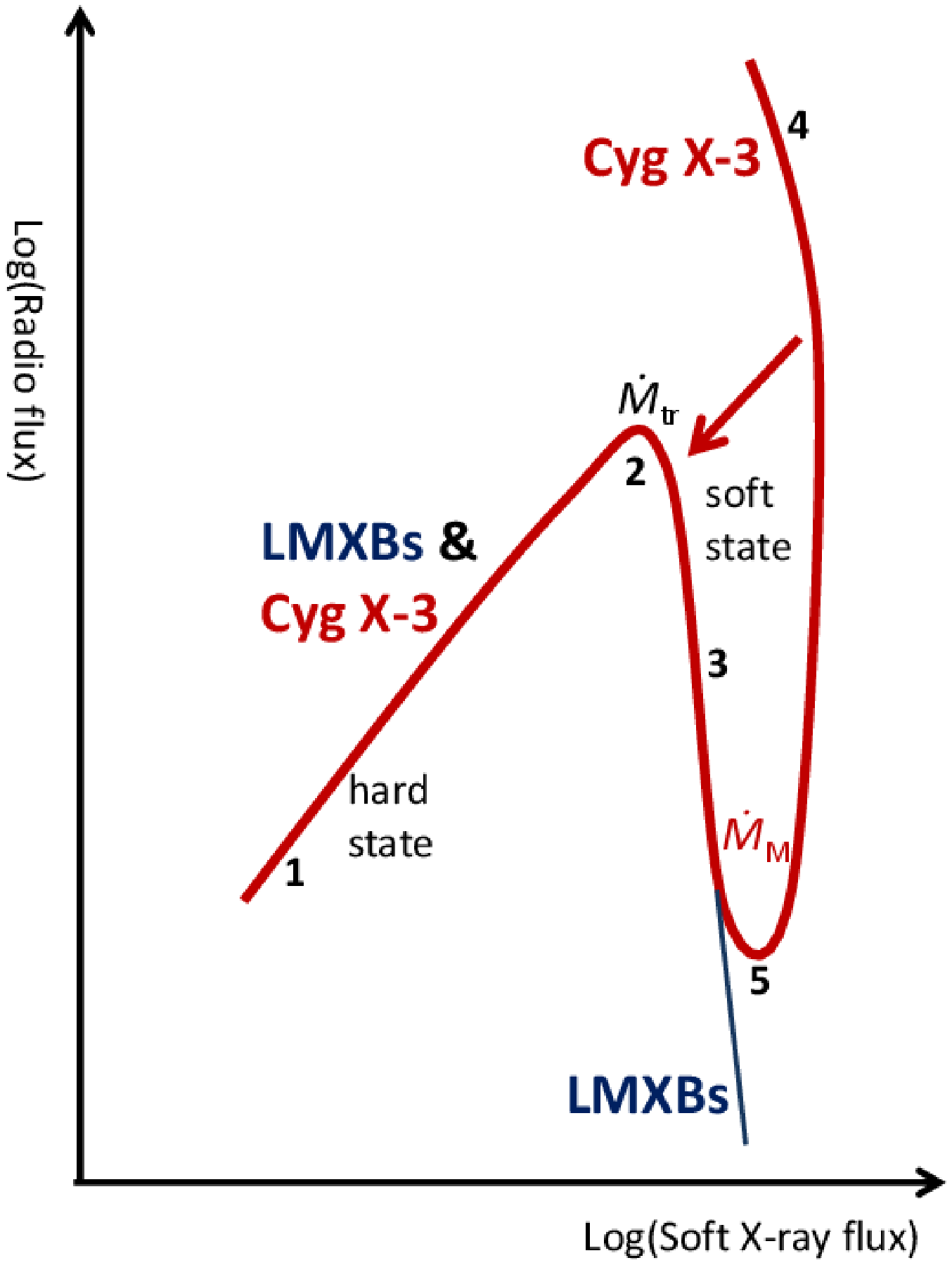}}
\caption{A schematic representation of the evolution of \source through its radio and soft X-ray states, as given by the radio flux vs.\ the soft X-ray flux measured on the same days. The diagram is based on the results of \citet{szm08}, now also showing the radio quenching of BH LMXBs in the soft state in the thin blue line. The evolution of Cyg X-3 occurs along the thick red curve, except that some transitions back from the high radio fluxes occur to the hard state, as shown by the arrow. Transitions between the hard and soft states occur at $\dot M_{\rm tr}$, and the onset of the magnetic outflows occurs at $\dot M_{\rm M}>\dot M_{\rm tr}$. The numbers correspond to the \source X-ray spectral states in the classification of \citet{szm08}, as shown in their fig.\ 2, with states 1 and 5 being the hardest and softest, respectively.  
\label{pattern} }
\end{figure}

The distance to \source has been estimated as $\approx$6--8\,kpc \citep*{lzt09,McCollough16,koljonen17}. The orbital period is $P\approx 0.20$\,d, see, e.g., \citet{bhargava17} for a recent reference. The masses of the binary components are quite uncertain. They were estimated by \citet{zmb13} based on the measurements of the radial velocities of \citet*{hanson00} and \citet{vilhu09}. However, those measurements bear relatively large systematic uncertainties (\citealt{koljonen17}; K. Koljonen, private communication). Still, both \citet{zmb13} and \citet{koljonen17} found the BH mass to be small. Hereafter, we scale our results to a fiducial BH mass of $M=5\msun$. The mass of the WR donor, $M_{\rm WR}$, appears to be $\sim\! 10\msun$ \citep{zmb13}. For its radius, we use the estimates of \citet{langer89}, which give $R_{\rm WR}\approx 0.9\rsun$ at this $M_{\rm WR}$. (Note that $R_{\rm WR}$ corresponds to the star itself, without including the wind, which is optically thick in \source; thus $R_{\rm WR}$ is significantly smaller than the photospheric radius.) 

Given strong X-ray absorption in \source and the current lack of reliable spectral models, its bolometric flux, $F$, is relatively uncertain. \citet*{z16} estimated $F$ in all of the states based on the requirement that there is the radio flux is positively correlated with the bolometric luminosity in the hard state. This requires a rather high absorbing column, and the average values of $F$ in the hard, soft and intermediate states were estimated as, respectively, 2.3, 5.5 and $8.0\times 10^{-8}$\,erg cm$^{-2}$ s$^{-1}$. On the other hand, \citet{kallman19} found weaker absorption, which would imply lower fluxes.  

The study of long-term variability of \source of \citet{z18} has shown that the radio flux is positively correlated with soft X-rays in the hard spectral state with any lag constrained to be $\ll$1\,d, and at a high Pearson's correlation coefficient of $r_{\rm P}\approx 0.7$, as shown in Fig.\ \ref{corr}(a). The radio emission is then suppressed after its transition to the soft state \citep[e.g.,][]{szm08}, as shown in Fig.\ \ref{pattern}. This behaviour is the same as that observed in BH LMXBs. However, after spending some time in the radio quenched state, strong radio flares (up to $\sim$20 Jy) appear in \source, but not in LMXBs, as shown in Fig.\ \ref{pattern}. The average time lag of the delayed radio flaring in \source has been measured \citep{z18} as $\approx$45--50\,d, appearing as a distinct peak in the cross-correlation, see Fig.\ \ref{corr}(b). We note that the peak is relatively wide, implying a correspondingly large dispersion of the lag in individual cases. The return from the major flaring state can proceed either back to the radio-quenched state or directly to the hard state, as indicated in Fig.\ \ref{pattern}. We stress that Fig.\ \ref{pattern} shows the evolution in a simplified way, while individual events  in \source follow varying trajectories, see Figs.\ 4 and 5 of \citet{szm08}. See also Fig.\ 7 of \citet{islam18} for the correlation of the radio flux with the bolometric flux, where their data show the highest radio fluxes corresponding to the decline of the soft state or even the hard state, consistent with their delayed emission.

\begin{figure}
\centerline{\includegraphics[width=7.cm]{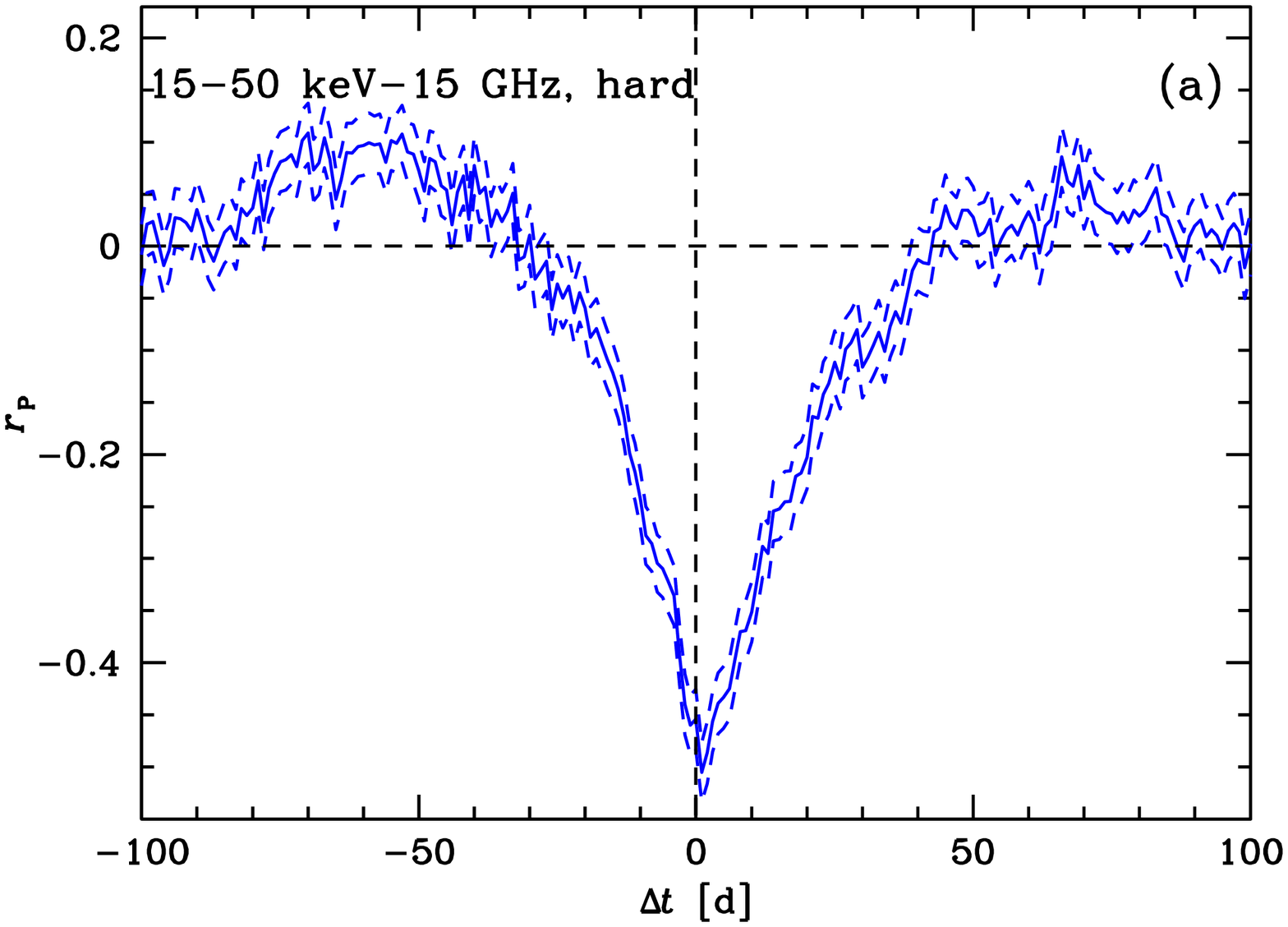}}
\centerline{\includegraphics[width=7.cm]{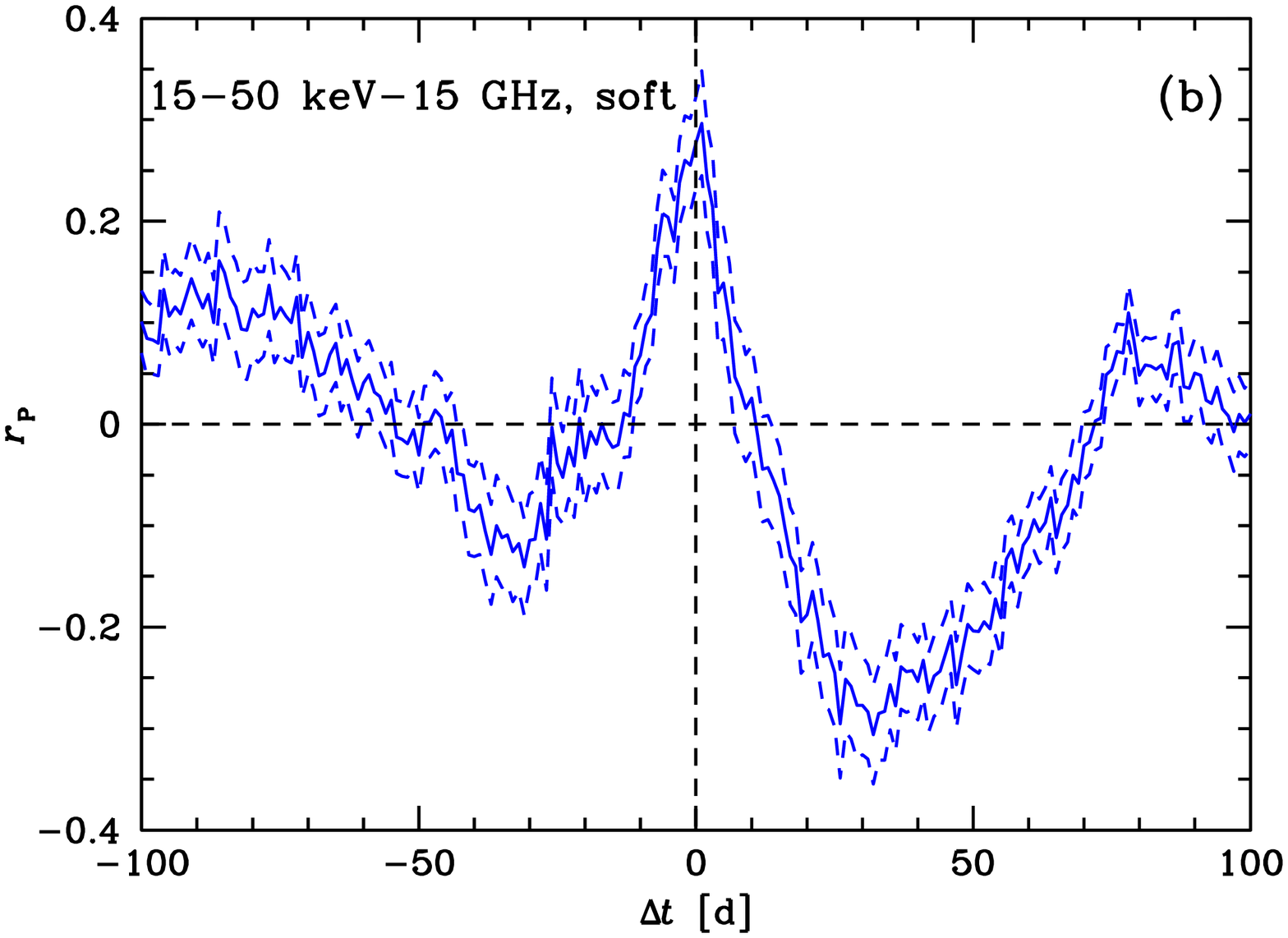}}
\caption{The cross-correlations (middle solid lines) between the 15--50 keV hard X-ray and the 15 GHz radio fluxes in the (a) hard and intermediate and (b) soft spectral states of \source. The dashed lines show the $1\sigma$ uncertainty ranges. The X-ray data are from the Burst Alert Detector \citep{krimm13} on board {\it Neil Gehrels Swift}. The radio data are from the Ryle and AMI telescopes. The hard/intermediate (soft) state is assumed to correspond to the 15--50 keV count rate of $>$0.028 ($<$0.025) cm$^{-2}$ s$^{-1}$. In addition, former is restricted to the range of the radio flux of 30--300 mJy. Adapted from \citet{z18}. 
\label{corr_bat} }
\end{figure}

Figs.\ \ref{corr_bat}(a,b) show the corresponding cross-correlations of the hard X-ray fluxes with radio in \source. In the hard state, we see a reversal of the positive zero-lag correlation with soft X-rays, which is due to the hard-state X-ray spectra pivoting around 10--15 keV. However, there is now a significant positive peak at a lag $\ll$1\,d in the soft state, different from $r_{\rm P}\approx 0$ at zero lag for soft X-rays. In addition, there is a significant anti-correlation at a radio lag of $\sim$30\,d.

\section{The Model}\label{model}

Unlike the case of BH LMXBs, strong relativistic jets are observed in the soft state of \source. Here, we propose that a strong magnetic field near the BH in this state, required for the jet formation, is formed by advection of the magnetic field from the companion WR star in a thin accretion disc, and the field advection is driven by magnetohydrodynamical (MHD) outflows. For brevity, we will hereafter use the term `magnetic outflows'. In this model, most of the angular momentum of the thin disc is removed by the outflows via the Blandford-Payne mechanism, which leads to high radial velocities and efficient field advection in the disc (see \citealt{2013ApJ...765..149C, 2019ApJ...872..149L} for details). This means the conventional viscous thin disc formed after the transition to the soft state will evolve into a thin disc with magnetic outflows. 

We mention here that there is another detailed model considering the disc-outflow system, in which the removal of angular momentum of the gas in the disc by MHD outflows has also been properly considered, developed by \citet{ferreira95} and \citet{ferreira06}. However, a rather simplified self-similar disc-outflow solution is derived in their calculations. Their model is further applied to explain the observational features in different states of XRBs, see, e.g., \citet{marcel18}.

It has been found that a minimal field strength is required to trigger magnetic outflows \citep{2019MNRAS.485.1916C}. The stellar wind launched by the star is threaded by its magnetic field. As the magnetic field flux is (approximately) frozen within the magnetized stellar wind, the field strength of the disc formed by the wind will follow the mass accretion rate. Thus, an increase of this rate may lead to the field strength being sufficiently high to trigger outflows, which will allow the field to be efficiently dragged inwards by the gas in the disc and enhanced \citep{2013ApJ...765..149C}. This will then allow formation of relativistic jets when the advected vertical flux approaches the BH. The observed (Fig.\ \ref{corr}b) time delay of the radio emission from the jets with respect to the soft X-ray emission (from the disc) corresponds then to the time between the disc reaching the innermost stable circular orbit (ISCO) during or after a hard-to-soft transition and the advected vertical flux reaching the BH. 

In the hard state of \source, the radio emission positively correlates with the soft X-ray emission, with no time lag detected (Fig.\ \ref{corr}a). As a hot accretion flow is present in the hard state, one may conjecture that the large-scale field is generated in situ in the hot flow (e.g., \citealt*{liska18}), which can naturally explain nearly simultaneous variability in the radio and X-ray bands.

\subsection{Transfer of gas and magnetic field}\label{tau_adv}

The magnetic induction equation is
\begin{equation}
{\frac {\partial {\bm B}}{\partial t}}=\triangledown\times{\bm v}\times{\bm B}-\triangledown\times(\eta\triangledown\times{\bm B}). \label{induct_0}
\end{equation}
For an axially symmetric accretion disc, this equation becomes
\begin{equation}
{\frac {\partial B_z}{\partial t}}={\frac 1{R}}{\frac {\partial}{\partial R}}[R(v_zB_r-v_rB_z)]-{\frac 1{R}}{\frac {\partial}{\partial R}}\left[R\eta\left({\frac {\partial B_r}{\partial z}}-{\frac {\partial B_z}{\partial R}} \right)\right],
\label{induct_1}
\end{equation}
where $R$ is the disc radius, $v_r$ and $v_z$ are the radial and vertical velocities of the disc, respectively. In our simple estimates below, we assume $v_z=0$, i.e., we neglect the effect of the outflow on the structure of the magnetic field. For a geometrically thin accretion disc, $\partial B_r/\partial z\thickapprox  B_r^{\rm S}/H$, where $B_r^{\rm S}$ is the radial component of the field at the disc surface and $H$ is the half-thickness of the disc. In the case of $B_r^{\rm S}\sim B_z$, i.e., the field line being significantly inclined with respect to the $z$ axis at the surface, we have $|\partial B_z/\partial R|\sim B_z/R\ll\partial B_r/\partial z\sim B_r^{\rm S}/H$ for a thin disc with $H/R\ll 1$. Thus, the induction equation (\ref{induct_1}) for a thin accretion disc can be rewritten as
\begin{equation}
{\frac {\partial B_z}{\partial t}}=-{\frac 1{R}}{\frac {\partial}{\partial R}}(Rv_rB_z)
-{\frac 1{R}}{\frac {\partial}{\partial R}}{\frac {\eta R B_r^{\rm S}}{H}}.
\label{induct_2}
\end{equation}

In order to explore the evolution of the magnetic field dragged inwards by the accretion disc, we would need, in principle, to solve the induction equation (\ref{induct_2}) together with a set of equations describing the disc and outflow structure with suitable initial and boundary conditions specified, which is beyond the scope of this work. Instead, we here simply estimate the timescale for the field flux co-moving with the gas from the outer edge of the disc to reach the BH, in the case of a disc with magnetic outflows.

Neither the detailed physics of the gas transfer from the companion star to the outer edge of the disc nor the magnetic field strength at the stellar surface are known. We only are reasonably certain that the wind is both enhanced around the binary plane and gravitationally focused toward the compact object. However, without loss of generality, we can estimate a scaling-law dependence of the field strength of the outer region of the disc with the mass accretion rate. Suppose the wind leaves the star from a certain region at the star surface in the direction perpendicular to the magnetic field. Then we have
\begin{equation}
\frac {B_z(R_{\rm d})}{\Sigma(R_{\rm d})}=f_{\rm B}\frac {B_*}{\Sigma_*},
\label{b_freeze}
\end{equation}
where $R_{\rm d}$ is the disc outer radius, $\Sigma$ is the disc column density,
$B_*$ is the field strength at the surface of the star, $\Sigma_*$ is the column density of the region on the stellar surface, and the efficiency of the field flux transport from the star to the disc is $f_{\rm B}\le 1$ ($f_{\rm B}=1$ for a perfectly conductive plasma without diffusion, i.e., for flux freezing). We assume that the gas density and geometry of this region remain unchanged when the mass accretion rate varies, i.e., the variation of mass accretion rate is predominantly caused by the change of the wind velocity. Then
\begin{equation}
B_z(R_{\rm d})\propto \Sigma(R_{\rm d})\propto \dot{M}^{7/10},
\label{b_mdot}
\end{equation}
where $\dot M$ is the mass accretion rate and $\Sigma\propto\dot{M}^{7/10}$ is adopted \citep*[see equation 5.49 in][]{frank02}. 

On the other hand, the condition
\begin{equation}
B_z(R_{\rm d})\ga B_z^{\rm min}\approx 2^{-\frac{1}{2}}\dot M^{\frac{1}{2}}\Omega_{\rm K}^{\frac{1}{2}} R^{-\frac{1}{2}}\approx
9.8\times 10^8{\rm G}\, \dot{m}^{\frac{1}{2}} m^{-\frac{1}{2}}r_{\rm d}^{-\frac{5}{4}} \label{bz_min1}
\end{equation}
has been found to be required for an accretion disc with magnetically driven outflows to exist. This follows from the condition of $B_\phi^{\rm S}/B_z\la 1$, required in order to avoid the instability/reconnection of toroidal fields \citep{2019MNRAS.485.1916C}. Here, $\Omega_{\rm K}\equiv (G M/R^3)^{1/2}$ is the Keplerian angular velocity, $B_\phi^{\rm S}$ is the toroidal component of the field at the disc surface, and the dimensionless mass accretion rate, $\dot m$, mass, $m$, and radius, $r$, are defined by
\begin{equation}
m\equiv {\frac {M}{M_\odot}},\quad \dot{m}\equiv \frac{0.1\dot M c^2}{L_{\rm
Edd,H}}= \frac {L}{\epsilon_{0.1}L_{\rm
Edd,H}}, \quad r\equiv {\frac {R}{R_{\rm g}}},\label{pars}
\end{equation}
respectively. Here $R_{\rm g}=GM/c^2$, $L$ is the bolometric luminosity of the accretion flow, $\epsilon_{0.1}$ is the accretion radiative efficiency in units of 0.1, and $L_{\rm Edd,H}$ is the Eddington luminosity for a pure H composition. Note that the donor of \source is probably a He star, and then the actual Eddington luminosity is twice higher. However, we use $\dot m$ defined above for consistency with other works.

From equation (\ref{bz_min1}), the required minimal field strength increases with the decreasing disc radius. This implies that the outflows would be driven from the outer region of the disc first. Most likely, the observed X-ray variability in \source is caused by variation of the mass accretion rate, with an increase of the mass supply to the disc leading to an increasing X-ray flux. The outflows may be triggered in the outer disc region only if the field strength surpasses the minimal value given by equation (\ref{bz_min1}), which is $\propto \dot M^{1/2}$. However, the actual value of the magnetic field strength increases faster, as $\propto \dot M^{7/10}$, see equation (\ref{b_mdot}). Thus, there will be {\it a minimum value of $\dot M$ above which the disc has magnetically driven outflows}, which are required for efficient advection of external magnetic field. We will denote it by $\dot M_{\rm M}$. In our case, $\dot M_{\rm M}> \dot M_{\rm tr}$ is required.

When the mass accretion rate reaches a peak value, $\partial B_z/\partial t\approx 0$ becomes a good approximation. Thus, we have
\begin{equation}
-v_{r}R B_z\approx\eta{\frac {R}{H}}B_r^{\rm S}
\label{induct_3}
\end{equation}
from equation (\ref{induct_2}). The field advection timescale in the case with magnetic outflows removing the angular momentum can then be estimated with
\begin{equation}
\tau_{\rm adv}(R) \equiv {\frac R{|v_r|}}={\frac {R H\kappa_0}{\eta}}=\frac{R_{\rm d} P_{\rm m} \kappa_0}{\alpha c_{\rm s} }
=\alpha^{-1}\left({\frac {H}{R}}\right)^{-1}\Omega_{\rm K}^{-1}P_{\rm m}\kappa_0, \label{tau_adv_dif}
\end{equation}
where $\kappa_0\equiv B_z/B_{r}^{\rm S}$ at the disc surface, $c_{\rm s}$ is the sound speed, and the viscosity scaling law of $\nu=\alpha c_{\rm s}H$, $c_{\rm s}=H\Omega_{\rm K}$ and equation (\ref{induct_3}) have been used. This timescale is shorter by $\sim\! H/R$ than the standard viscous time scale (i.e., in discs without magnetic outflows),
\begin{equation}
\tau_{\rm visc}(R) =\frac {R^2}{\alpha c_{\rm s}H} =\alpha^{-1}\left(\frac {H}{R}\right)^{-2}\Omega_{\rm K}^{-1}. 
\label{tau_visc}
\end{equation}

On the other hand, the shortest timescale in discs with magnetic outflows is that of the disc-outflow coupling, which is of the order of the disc dynamical timescale \citep{2002A&A...385..289C}, which is, in turn, negligibly small compared with the field advection timescale. Therefore, accretion in a disc with magnetic outflows proceeds at almost the same (high) speed as the field advection (with a small difference related to the diffusion timescale). We estimate the timescale required to accumulate sufficient magnetic field fluxes to drive jets from the region near the BH using the disc size estimates in Section \ref{parameters}.

The field inclination is crucial in accelerating outflows. The cold gas can be magnetically driven from the midplane of a Keplerian disc only if the field line is inclined with an angle $\le 60^\circ$ (i.e., $\kappa_0\le\sqrt{3}$) with respect to the midplane of the disc \citep{1982MNRAS.199..883B}. The situation is slightly different when the gas pressure plays an important role in magnetically driven outflows. In this, more realistic, case, the gas can be driven from the disc surface even if $\kappa_0$ is slightly higher than $\sqrt{3}$ \citep[see][]{1994A&A...287...80C,2001ApJ...553..158O,2002A&A...385..289C}. On the other hand, numerical simulations indicate that the Prandtl number is $P_{\rm m}\approx 1$--5 (see the references in Section \ref{intro}).

\subsection{Disc size and the time scales in \source}\label{parameters}

The main uncertainty in our estimates is the disc outer radius. In wind accreting binaries, it is usually much smaller \citep{sl76} than the tidally-limited value of $\sim\! 0.9 R_{\rm L}$ in Roche-lobe overflow binaries \citep{frank02}, where $R_{\rm L}$ is the Roche lobe radius of the compact object. However, \source is a very compact binary, with the orbital period of $P\approx 0.20$\,d, and therefore the Roche lobe radius is small. Also, the wind velocity from the WR donor appears to be rather low, $v_{\rm w}\approx 700$--800 km s$^{-1}$ at the location of the compact object, or even less if X-ray ionization substantially diminishes the wind acceleration on the irradiated side of the donor \citep{koljonen17}. We first try to estimate the disc radius using the formalism of \citet{hoyle39} as given in \citet{sl76}. The radius of an accretion cylinder (see fig.\ 1 in \citealt{sl76}) is $R_{\rm acc}\approx 2 G M v_{\rm rel}^{-2}$, where $v_{\rm rel}^2=v_{\rm w}^2+v_{\rm orb}^2$, $v_{\rm orb}$ is the relative orbital velocity of the two components. Usually, e.g., in Cyg X-1, $R_{\rm acc}\ll A$, where $A$ is the orbital separation. However, $A\approx 3.5[(M+M_{\rm WR})/15]^{1/3}\rsun$ and $R_{\rm acc}/A\approx (1.2$--$1.4)q/(1+q)$ in \source (or even more if the wind acceleration is reduced by ionization), which is typically larger than the size of the Roche lobe around the compact object, $R_{\rm L}/A\approx 0.46 [q/(1+q)]^{1/3}$ \citep{paczynski67}. Here $q\equiv M/M_{\rm WR}$. Then, $R_{\rm acc}\ga R_{\rm L}$ renders the application of the formalism of \citet{sl76} invalid and implies the resulting disc will have the size comparable to that of the Roche lobe.

In this case, the accretion disc will be still limited tidally, with $R_{\rm d}\la 0.9 R_{\rm L}$. For $q\la 2.5$, the Roche lobe radius estimate of \citet{paczynski67} gives a fractional error of $\la$10 per cent (as compared to the estimate of \citealt{eggleton83}, which is very accurate for any $q$). Using the Kepler law combined with the estimate of \citet{paczynski67}, we have
\begin{equation}
R_{\rm d}\la 0.9R_{\rm L}\approx 0.9(2 G M)^{\frac{1}{3}} (P/9\upi)^{\frac{2}{3}}\approx 9.6\times 10^4  R_{\rm g}(M/5\msun)^{-\frac{2}{3}}.
\label{Roche}
\end{equation}
For a total mass of $15\msun$, $R_{\rm d}\la 0.29 A$. For the assumed masses, the disc outer radius is $\ga\! 1.6\rsun$ from the donor surface (with $R_{\rm WR}\simeq 0.9\rsun$, see Section \ref{observation}).

The sound speed at the outer edge of the disc is usually in the regime dominated by the bound-free opacities and gas pressure. The midplane temperature could be estimated from the standard disc structure, as, e.g., in equation 5.49 of \citet{frank02}, which yields $T_{\rm d}\sim 10^4$ K. However, the disc in \source is embedded in the strong blackbody radiation of the donor, with $T_*\approx 10^5$\,K. Thus, $T_{\rm d}\sim T_*$ is likely. The sound speed is
\begin{equation}
c_{\rm s}=\sqrt{\frac{\Gamma k T_{\rm d}}{\mu m_{\rm p}}},
\label{sound}
\end{equation}
where $\Gamma\approx 5/3$ is the adiabatic index, $\mu$ is the mean molecular weight, which is within $\mu\approx 4/3$ for fully ionized He and 4 for neutral He (which is most likely the main element of the WR star in \source), and $m_{\rm p}$ is the proton mass. At $10^5$\,K, He will be predominantly fully ionized, which yields $H/R_{\rm d}\approx 0.03$. The field advection timescale in the presence of magnetic outflows, equation (\ref{tau_adv_dif}), then becomes
\begin{equation}
\tau_{\rm adv}\approx 4.4\,{\rm d} \frac{R_{\rm d}}{0.9R_{\rm L}}
\left(\frac{T_{\rm d}}{10^5\,{\rm K}}\right)^{-\frac{1}{2}} \left(\frac{\mu}{4/3}\right)^\frac{1}{2} \left(\frac{M}{5\msun}\right)^\frac{1}{3} \left(\frac{\alpha}{0.1}\right)^{-1}\! P_{\rm m} \frac{\kappa_0}{\sqrt{3}}.
\label{tau_adv2}
\end{equation}
On the other hand, the standard viscous time scale (i.e., in the absence of outflows), equation (\ref{tau_visc}), is
\begin{equation}
\tau_{\rm visc} \approx 77\,{\rm d} \left(\frac{R_{\rm d}}{0.9R_{\rm L}}\right)^\frac{1}{2}
\left(\frac{T_{\rm d}}{10^5\,{\rm K}}\right)^{-1} \frac{\mu}{4/3} \left(\frac{M}{5\msun}\right)^\frac{2}{3} \left(\frac{\alpha}{0.1}\right)^{-1}. 
\label{tau_visc2}
\end{equation}
We see that the latter timescale estimate is of the order of the average $\sim$50-d radio lag observed in \source. It may also be consistent with it for lower values of $R_{\rm d}/R_{\rm L}$ and some plausible combinations of the parameters. 

Equation (\ref{bz_min1}) estimates the value of the minimum strength of the magnetic field at $R_{\rm d}$ required for the appearance of magnetic outflows. Using our estimates of the parameters of \source, it can be written as
\begin{equation}
B_z^{\rm min}(R_{\rm d})\approx 80\, {\rm G}\left(\frac{F/\epsilon_{0.1}}{10^{-8}{\rm erg\,cm}^{-1} {\rm s}^{-1}}\right)^{\frac{1}{2}}\!\!\!\! \frac{d}{7\,{\rm kpc}}\left(\frac{R_{\rm d}}{0.9 R_{\rm L}}\right)^{-\frac{5}{4}} \!\left(\frac{M}{5\msun}\right)^{-\frac{1}{6}}.\label{bzmin}
\end{equation}
Current constraints give the average soft-state bolometric flux of $F\la 8.0\times 10^{38}$\,erg cm$^{-2}$ s$^{-1}$, as we discuss in Section \ref{observation}.  Likely values of $B_z^{\rm min}$ are then in the range of $\approx$100--200\,G. Note that the dependence on the uncertain BH mass is weak.

\subsection{The nature of the hard state}\label{hard_state}

In the low/hard state, a geometrically thick and hot inner accretion flow surrounded by an outer thin disc is present (e.g., see \citealt{2014ARA&A..52..529Y} and the references therein). The hot accretion flow may be advection dominated (hereafter abbreviated as ADAF) though the degree of the flow advection at high Eddington ratios is uncertain. Also, the flow is likely to contain some condensed cold clumps. Sill, for notational simplicity, we will use the term ADAF hereafter, but without implying a specific physics of the flow. Variations of the mass accretion rate trigger then state transitions. A hard-to-soft state transition occurs when the accretion rate increases above the critical value, $\dot M_{\rm tr}$. Then the ADAF cools efficiently and collapses to a thin accretion disc, which is accompanied by disappearance of jets.

Steady jets are universally observed in the hard state of BH XRBs, which implies the presence of strong vertical/poloidal magnetic fields near the BH. The origin of such strong fields is unclear. If the ADAF extended all the way to the outer edge of the disc in this state, any outer field could be efficiently advected on the viscous time scale \citep{2011ApJ...737...94C}. Equation (\ref{tau_visc}) then yields $\approx 0.3$\,d for an ADAF extending to $R_{\rm d}$ for typical values of $H/R_{\rm d}=0.5$ and $\alpha=0.1$. In principle, this could also explain the nearly simultaneous radio/soft X-ray variability in the hard state, see Fig.\ \ref{corr}(a).

However, it is highly unlikely that the ADAF is as large in \source, especially given its high Eddington ratio, $\ga$0.1 in the hard state. The exact ADAF extent at those Eddington ratios in the hard state in BH XRBs is a subject of a debate, but the disc truncation radii found by different methods are in the range of $\sim$2--$200 R_{\rm g}$, see, e.g., fig.\ 11 in \citet{garcia15}. Given those, the linear sizes of the outer thin disc are almost identical in both the hard and soft states. Some slow field advection in the outer thin viscous disc is possible \citep{1994MNRAS.267..235L,2019ApJ...872..149L} even below the threshold for magnetic outflows. Then the vertical field component transported to the truncation radius could be amplified in the ADAF, but, given its limited extent, it is unclear if this would be sufficient to explain the hard-state jets. Furthermore, the ADAF disappears in the soft state, but the field advection through the outer thin disc would still continue. This may be in conflict with the jet quenching during the hard-to-soft transitions, seen both in LXMBs and in \source (Fig.\ \ref{pattern}). 

On the other hand, it has been proposed that a large scale vertical field can be generated internally in thick discs (e.g., \citealt*{1996MNRAS.281..219T, 2004MNRAS.348..111K, 2014ApJ...782L..18B, parfrey15}). Indeed, recent MHD simulations indicate that accretion disc turbulence can generate large-scale poloidal magnetic flux in situ, even starting from a purely toroidal magnetic field, which may drive jets near the BH \citep{liska18}. Any variation of the mass accretion rate will affect the both the structure of the ADAF and the in-situ field generation, which would lead to nearly simultaneous variability in the radio and X-ray emission.

The field internally generated in the ADAF will decay after the transition to a thin accretion disc. Estimates of the state transition timescale and the field diffusion time scale in the thin disc show these two timescales are roughly comparable \citep{2016ApJ...817...71C, 2019MNRAS.485.1916C}, which implies that the jets will be soon switched off after the hard-to-soft state transition.

\subsection{The scenario for \source and LMXBs}
\label{scenario}

Based on our investigations above, we obtain the following physical explanations for the states of both \source and BH LMXBs.

In the hard state, we find it most likely that the large scale field is generated in situ in the inner hot accretion flow. Then, advection of the magnetic field of the donor plays little role. This is consistent with the relative universality of the radio/X-ray correlation in LMXBs (e.g., \citealt{corbel13}), since advection of the fields of the donors would lead to a range of magnetic fluxes, depending on the donor magnetic field, and, probably, to widely different jet radio fluxes. It is also consistent with the limited extent of the hot flow, which is surrounded by a thin disc, in which field advection is inefficient.

As the accretion rate increases and a transition to the soft state takes place at $\dot M_{\rm tr}$, the hot flow is replaced by a thin disc (without magnetic outflows), which both quenches the internal field generation and causes the field to quickly diffuse away. This leads to a disappearance of the jets and their radio emission, illustrated in Fig.\ \ref{pattern}. In \source, $\dot m_{\rm tr}\la 0.4$ can be estimated, see, e.g., fig.\ 7 in \citet{islam18}, where $L_{\rm bol}/L_{\rm Edd}\approx 0.2$, and $L_{\rm bol}$ is the bolometric accretion luminosity, $L_{\rm Edd}$ is the Eddington luminosity for an He-dominated composition (and this $L_{\rm bol}/L_{\rm Edd}$ corresponds to the absorbing column as estimated by \citealt{z16}). This stage corresponds to the radio-quenched state after the hard-to-soft transition, see Fig.\ \ref{pattern}. Possible origins of the residual radio emission of \source in the quenched state are discussed in \citet{koljonen18}. The X-ray spectrum in this state is consistent with almost pure disc blackbody emission, both in LMXBs and \source \citep{koljonen10,koljonen18}.

In our model, a further increase of $\dot M$ up to and above $\dot M_{\rm M}$ leads to $B_z$ exceeding is minimum value required for field advection, $B_z^{\rm min}$. This, in turn, leads to the appearance of magnetic outflows above the thin disc, which can efficiently advect the outer field and lead to formation of the powerful jets and the major radio flaring state. The advected field appears to be significantly stronger than that that internally generated, explaining much higher radio fluxes in \source reached in the soft state, up to $\sim$20 Jy, compared to the hard state, which have fluxes $\la$0.3 Jy (e.g., \citealt{z16}). In order to explain the absence of the major radio flaring state in LMXBs, the magnetic flux transferred from the WR donor in \source is required to be higher than those transferred from low-mass donors in LMXBs.

The observed $\Delta t\sim 50$\,d lag between the radio and soft X-ray fluxes then approximately corresponds to the time taken for $\dot M$ to increase from $\dot M_{\rm tr}$ to $\dot M_{\rm M}$. This time is limited viscously, and, as estimated in equation (\ref{tau_visc2}), it is of the order of the observed lag. At $\dot M_{\rm tr}$ is reached in the inner disc, the soft X-ray flux strongly increases. However, the radio flux increases from the quenched state to flaring only when $\dot M_{\rm M}$ is reached in the BH vicinity. The initial stage of an occurrence of the soft state gives the maximum lag, and as $\dot M$ increases (but still below $\dot M_{\rm M}$), the lag shortens. This dispersion reflects the relatively large width of the cross-correlation maximum, and the peak at $\approx$50 d corresponds to an average lag. Once the magnetic-outflow disc state is reached at $\dot M\geq \dot M_{\rm M}$, the field advection velocity is almost the same as the disc inflow velocity, due to their efficient coupling. Therefore, the mass inflow and field advection take the same time in this stage, and thus this stage does not contribute to the lag. Note that the radio fluxes observed at the same time as the highest soft X-ray fluxes correspond to the jet launched during some previous stages of the soft state (but still at $\dot M\geq \dot M_{\rm M}$). Obviously, $\dot M_{\rm M}>M_{\rm tr}$ in \source is required, see Fig.\ \ref{pattern}. Taking, in addition, the propagation time within the jet into account, we approximately have
\begin{equation}
\Delta t({\rm soft\,X\rightarrow radio})\approx \Delta t(\dot M_{\rm tr}\rightarrow \dot M_{\rm M})+\Delta t({\rm jet}).
\end{equation}
Given the measurements of the jet extent in the soft state, see the references in Section \ref{intro}, $\Delta t({\rm jet})$ it at most a few days, and probably much shorter. Thus, the main cause of the lag is that of the $\dot M$ increase between $\dot M_{\rm tr}$ and $\dot M_{\rm M}$ on the viscous timescale. On the other hand, $B_z^{\rm min}$ and thus $\dot M_{\rm M}$ are never reached in LXMBs due to the lower magnetic fluxes available from their donors.

As $\dot M$ decreases during and after a major flaring interval in \source, the evolution is back to either the hard state or the quenched soft state, see Fig.\ \ref{pattern}. The actual transition would then depend on the fractional decrease of $\dot M$. A return to the quenched state corresponds to a smaller decrease, but still below the critical $\dot M_{\rm M}$ for the outflow disc. The accumulated magnetic field would then quickly diffuse away. A return directly to the hard state corresponds to a larger decrease down to the hard state range, where the disc becomes truncated and internal field generation is efficient. In LMXBs, the return to the hard state is from their quenched soft state only, though at luminosities lower than for the hard-to-soft transitions, e.g., \citet{fender04}. 

We then consider the origin of the hard X-ray emission in \source. In the hard spectral state, both soft and hard X-rays originate in the hot flow. The anti-correlation of radio and soft X-rays at a short, $\ll$1\,d, lag, shown in Fig.\ \ref{corr_bat}(a), opposite to the positive correlation shown in Fig.\ \ref{corr}(a), is due to the spectrum within the hard state pivoting around $\sim$10--15\,keV (e.g., \citealt{szm08}). As we expect the bolometric accretion flux to be positively correlated with the radio flux, the soft X-rays appear to be significantly absorbed, in order for their flux to dominate the bolometric flux \citep{z16}. In LMXBs, both soft and hard X-rays originate in the hot accretion flow, the same as in \source. However, both are positively correlated with the radio emission, e.g., in GX 339--4 \citep{corbel03}, which is due to the pivot energy being at an energy $\gg$10\,keV.

Then, in the soft spectral state of \source, the soft X-rays are from the disc emission, but the hard X-rays are, most likely, from emission of a corona above the disc. Fig.\ \ref{corr_bat}(b) shows that the hard X-rays are positively correlated at a short, $\ll$1\,d, lag with the radio emission, which, in turn, lags the soft X-rays. This is consistent with the results of \citet{koljonen10}, who have shown that major radio flares in \source occur during the transition from the supersoft state to harder ones, i.e., are associated with the appearance of a hard X-ray tail beyond the disc blackbody. Both radio and hard X-rays appear to be an effect of the magnetic outflows reaching the ISCO, and the hard X-ray emitting corona, possibly outflowing, is magnetically generated above an inner part of the accretion disc. However, it is uncertain whether the jet is produced directly by the corona. The short lag of radio with respect to hard X-rays also implies a fast signal propagation between the hard X-ray and radio emitting regions. On the other hand, there is no detectable radio flux in the soft state of LMXBs, as well as the hard X-ray emission (at energies above those of the disc blackbody) is very weak, consistent with the absence of magnetic field advection in the soft states of those sources.

Then, the anti-correlation seen at a $\sim$30-d radio lag with respect to hard X-rays in \source, Fig.\ \ref{corr_bat}(b), appears to reflect the typical duration of a radio flaring event. Namely, an increasing flux of the magnetic corona leading to radio flare with a short lag will be reflected by a declining radio flux $\sim$30-d later, when the radio flare disappears. 

\section{Discussion}\label{discuss}

Our model explains the unique feature of \source among BH XRBs in having powerful jets in the soft spectral state by postulating that its WR donor possesses strong magnetic field, which is advected to the BH above certain threshold accretion rate, $\dot M_{\rm M}$. On the other hand, we postulate that the magnetic fluxes transferred from the donors in LMXBs are lower and this effect does not take place in their soft states. 

\source has some other unique characteristics. Its hard-state X-ray spectra have high-energy breaks at lower energies and its power spectra are cut off above a much lower frequency than other BH XRBs, see, e.g., \citet{szm08} and \citet*{alh09}, respectively. These two characteristics have been explained by the presence of a Thomson-thick bulge around the central BH formed by the very strong WR wind. Then Compton scattering in the bulge both down-scatters the hard X-rays and removes high frequencies from the power spectra \citep*{zmg10}. \source is also unique among accreting X-ray binaries in its very strong high-energy $\gamma$-ray emission, associated with the major radio flares. As discussed, e.g., in \citet{z12}, this appears to be due to the very strong blackbody flux from the bright donor ($L_{\rm WR}\sim 10^{39}$\,erg\,s$^{-1}$) at a small orbital separation from the BH. This results in the dominance of inverse Compton losses and $\gamma$-ray emission by relativistic electrons in the jets. On the other hand, BH LMXBs both do not possess such powerful jets and the blackbody fluxes from their donors are very weak. The weak blackbody flux is then likely to result in synchrotron emission dominating the electron energy losses. Altogether, those results explain most of the unique features of \source and provide a framework unifying it with LMXBs.

Details of our model are sensitive to the viscosity parameter, with both the viscous and magnetic timescales $\propto \alpha^{-1}$. The value of $\alpha$ is still not well constrained by either observations or theory. In dwarf novae, $\alpha\approx 0.1$--0.2 was found in the hot discs (e.g., \citealt{1999AcA....49..391S, 2012A&A...545A.115K}). Numerical simulations give $\alpha\approx 0.05$--0.2 \citep{2002ApJ...573..738H, 2013MNRAS.428.2255P} or $\alpha\approx 0.08$ for weak field cases, while $\alpha\ga 1$ for moderate field strength \citep{2013ApJ...767...30B}. \citet{2014ApJ...782L..18B} propose that $\alpha\approx 0.01$--0.1 in the soft state of BH XRBs in a model explaining the hysteretic cycle of state transitions. \citet{1997ApJ...489..865E} found $\alpha=0.25$ when modelling state transitions of BH XRBs. A Bayesian analysis of X-ray light curves of 21 outbursts of BH LMXBs with an accretion model yield $\alpha\approx 0.2$--1.0 \citep{2018Natur.554...69T, 2018MNRAS.480....2T}. For \source, no constraint on $\alpha$ is available. 

Considering the viscous time, equation (\ref{tau_visc2}), which we have identified as responsible for the radio vs.\ soft X-ray lag, we see that using a large value of $\alpha$ allows for values of the outer disc radius lower than its tidal truncation value. This, however, would increase the required minimum value of the vertical magnetic field at the disc outer edge to high values as $B_z^{\rm min}\propto R_{\rm d}^{-5/4}$, equation (\ref{bzmin}), which appears unlikely. Uncertainties in other parameters in those equations prevent more precise estimates. Furthermore, these parameters likely differ from one to another occurrence of the soft state associated with radio flaring.

In our model, the formation of a powerful jet at a large accretion rate is due to triggering magnetic outflows above certain critical value, $\dot M_{\rm M}$, which results in a rapid advection of the vertical magnetic flux from the donor. An alternative way to advect the fields is by postulating the presence of a hot coronal gas above/below the thin disc \citep{2009ApJ...707..428B, 2009ApJ...701..885L, 2012MNRAS.424.2097G,2013MNRAS.430..822G}. As the dynamics of a hot corona is similar to that of an ADAF, its field advection timescale is of the same order as that of the ADAF, i.e., a fraction of a day. While this timescale by itself cannot explain the observations, the lag can be due to the time spent in the soft state before the appearance of the corona, similarly to our magnetic-outflow model. However, it is not clear in which way an increase of the accretion rate above a certain value in the soft state would lead to the onset of coronal accretion starting at the disc outer edge. 

Another alternative explanation for the radio lag observed in \source is propagation within the jets. Since the resolved radio structures in the soft state are observed up to a distance of $\sim$1 light day \citep{2001ApJ...553..766M, 2004ApJ...600..368M, 2017MNRAS.471.2703E}, the required propagation speed is $\la$2 per cent of $c$, which does not appear likely, but it can still be the case. However, this would not explain the nearly simultaneous variability of radio and hard X-rays. Also, we would be still left with the issue why powerful radio jets are observed in \source but not in the soft state of LMXBs. 

\section{Conclusions}
\label{conclusions}

Our main conclusions are as follows.

We have carried out calculations of the process of advection of outer vertical magnetic field through a thin accretion disc. Our main finding is that there is a threshold mass accretion rate, $\dot M_{\rm M}$, above which the disc acquires magnetic outflows starting from its outer edge and is able to quickly advect the field to the BH. We postulate that this rate in \source is higher than the mass accretion rate for the hard-to-soft transition, $\dot M_{\rm tr}$. Thus, the transition is first associated with the quenching of the jets and their radio emission, as in BH LMXBs. Only when $\dot M$ further increases above $\dot M_{\rm M}$, the disc is able to advect the field in its magnetic outflow state. The observed long radio lag of $\sim$50\,d with respect to the soft X-rays is due, on average, to the viscous time scale of the accretion rate increase between $\dot M_{\rm tr}$ and $\dot M_{\rm M}$. 

The threshold for the magnetic outflows corresponds to a certain value of the vertical magnetic field at the disc outer edge. Only when such field is present this process happens. That field required in \source is $\sim$100--200 G, which we postulate is advected from the donor in the stellar wind. Thus, the Wolf-Rayet field strength has to be sufficiently high, though we have not been able to estimate it accurately. The difference between \source and LMXBs is postulated to be due to the magnetic fluxes available through Roche-lobe overflow from the donors of the latter being relatively low, leading to the values of the magnetic field below the threshold values.

We have calculated the size of the disc in \source and found the theoretical wind accretion radius to be larger than the size of the Roche lobe, which is due to both the low wind velocity and small binary separation. This implies that the actual disc size is comparable to that of the Roche lobe. This both reduces the value of the required threshold field for the appearance of magnetic outflows and facilitates the transfer of the magnetic field from the companion. 

We then consider the origin of the hard X-ray emission in the soft state. It is correlated with the radio at a $\ll$1-d lag. We explain the hard X-ray emission by the magnetic outflows producing a magnetic corona close to the BH. The observed short lag between hard X-rays and radio also implies a fast signal propagation along the jet up to the radio-emitting region.

We have also pointed out that outer cold discs in both \source and LMXBs are quite similar in both soft and luminous hard states. Thus, the common assumption that the jet quenching in the soft state is due to the inability of the thin disc to advect outer field cannot explain the presence of jets and radio emission in the hard state, which has outer disc similar to that in the soft state. Instead, we find that the likely possibility is generation of the required field in situ, as demonstrated in recent MHD calculations. 

\section*{Acknowledgments}

We thank Karri Koljonen and Marek Sikora for valuable discussions and comments, and the referee for valuable comments. We acknowledge support from the European Union's Horizon 2020 research and innovation programme under the Marie Sk{\l}odowska-Curie grant agreement No.\ 798726, the Polish National Science Centre under the grant 2015/18/A/ST9/00746, the NSFC grants 11773050 and 11833007, and the CAS grant QYZDJ-SSWSYS023.

\label{lastpage}


\begin{thebibliography}{}

\bibitem[\protect\citeauthoryear{Axelsson, Larsson \& Hjalmarsdotter}{Axelsson et al.}{2009}]{alh09} 
Axelsson M., Larsson S., Hjalmarsdotter L., 2009, MNRAS, 394, 1544

\bibitem[\protect\citeauthoryear{Bai \& Stone}{2013}]{2013ApJ...767...30B} Bai X.-N., Stone J.~M., 2013, ApJ, 767, 30

\bibitem[\protect\citeauthoryear{Beckwith, Hawley \& Krolik}{Beckwith et al.}{2009}]{2009ApJ...707..428B} Beckwith K., Hawley J.~F., Krolik J.~H., 2009, ApJ, 707, 428

\bibitem[\protect\citeauthoryear{Begelman \& Armitage}{2014}]{2014ApJ...782L..18B} Begelman M.~C., Armitage P.~J., 2014, ApJ, 782, L18

\bibitem[\protect\citeauthoryear{Belloni}{2010}]{2010LNP...794...53B} Belloni T.~M., 2010, in Lecture Notes in Physics, Vol. 794, The Jet
Paradigm. Springer-Verlag, Berlin, p.\ 53

\bibitem[\protect\citeauthoryear{Belloni et al.}{2005}]{2005A&A...440..207B} Belloni T., Homan J., Casella P., van der Klis M., Nespoli E., Lewin W.~H.~G., Miller J.~M., M{\'e}ndez M., 2005, A\&A, 440, 207

\bibitem[\protect\citeauthoryear{Bhargava et al.}{2017}]{bhargava17} Bhargava Y., et al., 2017, ApJ, 849, 141 

\bibitem[\protect\citeauthoryear{Bisnovatyi-Kogan \& Ruzmaikin}{1974}]{1974Ap&SS..28...45B} Bisnovatyi-Kogan G.~S., Ruzmaikin A.~A., 1974, Ap\&SS, 28, 45

\bibitem[\protect\citeauthoryear{Bisnovatyi-Kogan \& Ruzmaikin}{1976}]{1976Ap&SS..42..401B} Bisnovatyi-Kogan G.~S., Ruzmaikin A.~A., 1976, Ap\&SS, 42, 401

\bibitem[\protect\citeauthoryear{Blandford \& Payne}{1982}]{1982MNRAS.199..883B} Blandford R.~D., Payne D.~G., 1982, MNRAS, 199, 883

\bibitem[\protect\citeauthoryear{Blandford \& Znajek}{1977}]{1977MNRAS.179..433B} Blandford R.~D., Znajek R.~L., 1977, MNRAS, 179, 433

\bibitem[\protect\citeauthoryear{Bradt, Rothschild \& Swank}{Bradt et al.}{1993}]{brs93}
Bradt H. V., Rothschild R. E., Swank J. H., 1993, A\&AS, 97, 355

\bibitem[\protect\citeauthoryear{Cao}{2011}]{2011ApJ...737...94C} Cao X., 2011, ApJ, 737, 94

\bibitem[\protect\citeauthoryear{Cao}{2016a}]{2016ApJ...817...71C} Cao X., 2016a, ApJ, 817, 71

\bibitem[\protect\citeauthoryear{Cao}{2016b}]{2016ApJ...833...30C} Cao X., 2016b, ApJ, 833, 30

\bibitem[\protect\citeauthoryear{Cao}{2018}]{2018MNRAS.473.4268C} Cao X., 2018, MNRAS, 473, 4268

\bibitem[\protect\citeauthoryear{Cao \& Lai}{2019}]{2019MNRAS.485.1916C} Cao X., Lai D., 2019, MNRAS, 485, 1916

\bibitem[\protect\citeauthoryear{Cao \& Spruit}{1994}]{1994A&A...287...80C} Cao X., Spruit H.~C., 1994, A\&A, 287, 80
\bibitem[\protect\citeauthoryear{Cao \& Spruit}{2002}]{2002A&A...385..289C} Cao X., Spruit H.~C., 2002, A\&A, 385, 289
\bibitem[\protect\citeauthoryear{Cao \& Spruit}{2013}]{2013ApJ...765..149C} Cao X., Spruit H.~C., 2013, ApJ, 765, 149

\bibitem[\protect\citeauthoryear{Corbel et al.}{2001}]{corbel01} Corbel S., et al., 2001, ApJ, 554, 43

\bibitem[\protect\citeauthoryear{Corbel et al.}{2003}]{corbel03} Corbel S., Nowak M.~A., Fender R.~P., Tzioumis A.~K., Markoff S., 2003, A\&A, 400, 1007 

\bibitem[\protect\citeauthoryear{Corbel et al.}{2013}]{corbel13} 
Corbel S., Coriat M., Brocksopp C., Tzioumis A.~K., Fender R.~P., Tomsick J.~A., Buxton M.~M., Bailyn C.~D., 2013, MNRAS, 428, 2500 

\bibitem[\protect\citeauthoryear{de la Chevroti{\`e}re et al.}{2014}]{2014ApJ...781...73D} de la Chevroti{\`e}re A., St-Louis N., Moffat A.~F.~J., MiMeS Collaboration, 2014, ApJ, 781, 73

\bibitem[\protect\citeauthoryear{Eggleton}{1983}]{eggleton83} Eggleton P.~P., 1983, ApJ, 268, 368

\bibitem[\protect\citeauthoryear{Egron et al.}{2017}]{2017MNRAS.471.2703E} Egron E., et al., 2017, MNRAS, 471, 2703

\bibitem[\protect\citeauthoryear{Esin, McClintock \& Narayan}{Esin et al.}{1997}]{1997ApJ...489..865E} Esin A.~A., McClintock J.~E., Narayan R., 1997, ApJ, 489, 865

\bibitem[\protect\citeauthoryear{Fender et al.}{1999}]{1999ApJ...519L.165F} Fender R., et al., 1999, ApJ, 519, L165

\bibitem[\protect\citeauthoryear{Fender, Belloni \& Gallo}{Fender et al.}{2004}]{fender04} Fender R.~P., Belloni T.~M., Gallo E., 2004, MNRAS, 355, 1105

\bibitem[\protect\citeauthoryear{Fender, Gallo \& Jonker}{Fender et al.}{2003}]{2003MNRAS.343L..99F} Fender R.~P., Gallo E., Jonker P.~G., 2003, MNRAS, 343, L99

\bibitem[\protect\citeauthoryear{Fender, Hanson \& Pooley}{Fender et al.}{1999}]{1999MNRAS.308..473F} Fender R.~P., Hanson M.~M., Pooley G.~G., 1999, MNRAS, 308, 473

\bibitem[\protect\citeauthoryear{Ferreira \& Pelletier}{1995}]{ferreira95} Ferreira J., Pelletier G., 1995, A\&A, 295, 807

\bibitem[\protect\citeauthoryear{Ferreira et al.}{2006}]{ferreira06} Ferreira J., Petrucci P.-O., Henri G., Saug{\'e} L., Pelletier G., 2006, A\&A, 447, 813

\bibitem[\protect\citeauthoryear{Frank, King \& Raine}{Frank et al.}{2002}]{frank02} Frank J., King A., Raine D.~J., 2002, Accretion Power in Astrophysics. Cambridge University Press

\bibitem[\protect\citeauthoryear{Fromang \& Stone}{2009}]{2009A&A...507...19F} Fromang S., Stone J.~M., 2009, A\&A, 507, 19

\bibitem[\protect\citeauthoryear{Gallo, Fender \& Pooley}{Gallo et al.}{2003}]{2003MNRAS.344...60G} Gallo E., Fender R.~P., Pooley G.~G., 2003, MNRAS, 344, 60

\bibitem[\protect\citeauthoryear{Garc{\'{\i}}a et al.}{2015}]{garcia15} 
Garc{\'{\i}}a J.~A., Steiner J.~F., McClintock J.~E., Remillard R.~A., Grinberg 
V., Dauser T., 2015, ApJ, 813, 84

\bibitem[\protect\citeauthoryear{Guan \& Gammie}{2009}]{2009ApJ...697.1901G} Guan X., Gammie C.~F., 2009, ApJ, 697, 1901

\bibitem[\protect\citeauthoryear{Guilet \& Ogilvie}{2012}]{2012MNRAS.424.2097G} Guilet J., Ogilvie G.~I., 2012, MNRAS, 424, 2097

\bibitem[\protect\citeauthoryear{Guilet \& Ogilvie}{2013}]{2013MNRAS.430..822G} Guilet J., Ogilvie G.~I., 2013, MNRAS, 430, 822

\bibitem[\protect\citeauthoryear{Hanson, Still \& Fender}{Hanson et al.}{2000}]{hanson00} Hanson M.~M., Still M.~D., Fender R.~P., 2000, ApJ, 541, 308

\bibitem[\protect\citeauthoryear{Hawley \& Balbus}{2002}]{2002ApJ...573..738H} Hawley J.~F., Balbus S.~A., 2002, ApJ, 573, 738

\bibitem[\protect\citeauthoryear{Hoyle \& Lyttleton}{1939}]{hoyle39} Hoyle F., Lyttleton R.~A., 1939, PCPS, 35, 405

\bibitem[\protect\citeauthoryear{Hubrig et al.}{2016}]{2016MNRAS.458.3381H} Hubrig S., Scholz K., Hamann W.-R., Sch{\"o}ller M., Ignace R., Ilyin I., Gayley K.~G., Oskinova L.~M., 2016, MNRAS, 458, 3381

\bibitem[\protect\citeauthoryear{Islam \& Zdziarski}{2018}]{islam18} Islam N., Zdziarski A.~A., 2018, MNRAS, 481, 4513 

\bibitem[\protect\citeauthoryear{Kallman et al.}{2019}]{kallman19} Kallman T., et al., 2019, ApJ, 874, 51 

\bibitem[\protect\citeauthoryear{King et al.}{2004}]{2004MNRAS.348..111K} King A.~R., Pringle J.~E., West R.~G., Livio M., 2004, MNRAS, 348, 111

\bibitem[\protect\citeauthoryear{Koljonen \& Maccarone}{2017}]{koljonen17} Koljonen K.~I.~I., Maccarone T.~J., 2017, MNRAS, 472, 2181

\bibitem[\protect\citeauthoryear{Koljonen et al.}{2010}]{koljonen10} 
Koljonen K.~I.~I., Hannikainen D.~C., McCollough M.~L., Pooley G.~G., Trushkin S.~A., 2010, MNRAS, 406, 307 

\bibitem[\protect\citeauthoryear{Koljonen et al.}{2018}]{koljonen18} Koljonen K.~I.~I., Maccarone T., McCollough M.~L., Gurwell M., Trushkin S.~A., Pooley G.~G., Piano G., Tavani M., 2018, A\&A, 612, A27 

\bibitem[\protect\citeauthoryear{Kotko \& Lasota}{2012}]{2012A&A...545A.115K} Kotko I., Lasota J.-P., 2012, A\&A, 545, A115

\bibitem[\protect\citeauthoryear{Krimm et al.}{2013}]{krimm13} 
Krimm H.~A., et al., 2013, ApJS, 209, 14 

\bibitem[\protect\citeauthoryear{Langer}{1989}]{langer89} Langer N., 1989, A\&A, 210, 93 

\bibitem[\protect\citeauthoryear{Lesur \& Longaretti}{2009}]{2009A&A...504..309L} Lesur G., Longaretti P.-Y., 2009, A\&A, 504, 309

\bibitem[\protect\citeauthoryear{Li \& Cao}{2019}]{2019ApJ...872..149L} Li J., Cao X., 2019, ApJ, 872, 149

\bibitem[\protect\citeauthoryear{Ling, Zhang \& Tang}{Ling et al.}{2009}]{lzt09} 
Ling Z., Zhang S. \& Tang S., 2009, ApJ, 695, 1111

\bibitem[\protect\citeauthoryear{Liska, Tchekhovskoy \& Quataert}{Liska et al.}{2018}]{liska18} Liska M.~T.~P., Tchekhovskoy A., Quataert E., 2018, arXiv:1809.04608

\bibitem[\protect\citeauthoryear{Lovelace, Rothstein \& Bisnovatyi-Kogan}{Lovelace et al.}{2009}]{2009ApJ...701..885L} Lovelace R.~V.~E., Rothstein D.~M., Bisnovatyi-Kogan G.~S., 2009, ApJ, 701, 885

\bibitem[\protect\citeauthoryear{Lubow, Papaloizou \& Pringle}{Lubow et al.}{1994}]{1994MNRAS.267..235L} Lubow S.~H., Papaloizou J.~C.~B., Pringle J.~E., 1994, MNRAS, 267, 235

\bibitem[\protect\citeauthoryear{Marcel et al.}{2018}]{marcel18} Marcel G., et al., 2018, A\&A, 615, A57

\bibitem[\protect\citeauthoryear{McCollough, Corrales \& Dunham}{McCollough et al.}{2016}]{McCollough16} McCollough M.~L., Corrales L., Dunham M.~M., 2016, ApJ, 830, L36 

\bibitem[\protect\citeauthoryear{Miller-Jones et al.}{2004}]{2004ApJ...600..368M} Miller-Jones J.~C.~A., Blundell K.~M., Rupen M.~P., Mioduszewski A.~J., Duffy P., Beasley A.~J., 2004, ApJ, 600, 368

\bibitem[\protect\citeauthoryear{Mioduszewski et al.}{2001}]{2001ApJ...553..766M} Mioduszewski A.~J., Rupen M.~P., Hjellming R.~M., Pooley G.~G., Waltman E.~B., 2001, ApJ, 553, 766

\bibitem[\protect\citeauthoryear{Miyamoto et al.}{1995}]{1995ApJ...442L..13M} Miyamoto S., Kitamoto S., Hayashida K., Egoshi W., 1995, ApJ, 442, L13

\bibitem[\protect\citeauthoryear{Ogilvie \& Livio}{2001}]{2001ApJ...553..158O} Ogilvie G.~I., Livio M., 2001, ApJ, 553, 158

\bibitem[\protect\citeauthoryear{Paczy\'nski}{1967}]{paczynski67}
Paczy\'nski B., 1967, Acta Astron., 17, 287

\bibitem[\protect\citeauthoryear{Parfrey, Giannios \& Beloborodov}{Parfrey et al.}{2015}]{parfrey15} Parfrey K., Giannios D., Beloborodov A.~M., 2015, MNRAS, 446, L61 

\bibitem[\protect\citeauthoryear{Parker}{1979}]{1979cmft.book.....P} Parker E.~N., 1979, in Cosmical magnetic fields: Their origin and their activity. Oxford University Press

\bibitem[\protect\citeauthoryear{Penna et al.}{2013}]{2013MNRAS.428.2255P} Penna R.~F., S{\c a}dowski A., Kulkarni A.~K., Narayan R., 2013, MNRAS, 428, 2255

\bibitem[\protect\citeauthoryear{Pooley \& Fender}{1997}]{pf97} 
Pooley G.~G., Fender R.~P., 1997, MNRAS, 292, 925

\bibitem[\protect\citeauthoryear{S{\c{a}}dowski}{2016}]{sadowski16} S{\c{a}}dowski A., 2016, \mnras, 462, 960

\bibitem[\protect\citeauthoryear{Salvesen et al.}{2016}]{2016MNRAS.460.3488S} Salvesen G., Armitage P.~J., Simon J.~B., Begelman M.~C., 2016, MNRAS, 460, 3488

\bibitem[\protect\citeauthoryear{Shapiro \& Lightman}{1976}]{sl76}
Shapiro S.~L., Lightman A.~P., 1976, ApJ, 204, 555

\bibitem[\protect\citeauthoryear{Smak}{1999}]{1999AcA....49..391S} Smak J., 1999, Acta Astron., 49, 391

\bibitem[\protect\citeauthoryear{Spruit \& Uzdensky}{2005}]{2005ApJ...629..960S} Spruit H.~C., Uzdensky D.~A., 2005, ApJ, 629, 960

\bibitem[\protect\citeauthoryear{Szostek, Zdziarski \& McCollough}{Szostek et al.}{2008}]{szm08} Szostek A., Zdziarski A.~A., McCollough M.~L., 2008, MNRAS, 388, 1001

\bibitem[\protect\citeauthoryear{Tetarenko et al.}{2018a}]{2018MNRAS.480....2T} Tetarenko B.~E., Dubus G., Lasota J.-P., Heinke C.~O., Sivakoff G.~R., 2018a, MNRAS, 480, 2

\bibitem[\protect\citeauthoryear{Tetarenko et al.}{2018b}]{2018Natur.554...69T} Tetarenko B.~E., Lasota J.-P., Heinke C.~O., Dubus G., Sivakoff G.~R., 2018b, Nature, 554, 69

\bibitem[\protect\citeauthoryear{Tout \& Pringle}{1996}]{1996MNRAS.281..219T} Tout C.~A., Pringle J.~E., 1996, MNRAS, 281, 219

\bibitem[\protect\citeauthoryear{van Kerkwijk et al.}{1992}]{1992Natur.355..703V} van Kerkwijk M.~H., et al., 1992, Nature, 355, 703

\bibitem[\protect\citeauthoryear{van Kerkwijk et al.}{1996}]{1996A&A...314..521V} van Kerkwijk M.~H., Geballe T.~R., King D.~L., van der Klis M., van Paradijs J., 1996, A\&A, 314, 521

\bibitem[\protect\citeauthoryear{Vilhu et al.}{2009}]{vilhu09}
Vilhu O., Hakala P., Hannikainen D.~C., McCollough M., Koljonen K., 2009, A\&A, 501, 679

\bibitem[\protect\citeauthoryear{Yousef, Brandenburg \& R{\"u}diger}{Yousef et al.}{2003}]{2003A&A...411..321Y} Yousef T.~A., Brandenburg A., R{\"u}diger G., 2003, A\&A, 411, 321

\bibitem[\protect\citeauthoryear{Yu \& Yan}{2009}]{2009ApJ...701.1940Y} Yu W., Yan Z., 2009, ApJ, 701, 1940

\bibitem[\protect\citeauthoryear{Yuan \& Narayan}{2014}]{2014ARA&A..52..529Y} Yuan F., Narayan R., 2014, ARA\&A, 52, 529

\bibitem[\protect\citeauthoryear{Zdziarski \& Gierli\'nski}{2004}]{zg04}
Zdziarski A.~A., Gierli\'nski M., 2004, Progr.\ Theor.\ Phys.\ Suppl., 155, 99

\bibitem[\protect\citeauthoryear{Zdziarski, Misra \& Gierli{\'n}ski}{Zdziarski et al.}{2010}]{zmg10} 
Zdziarski A.~A., Misra R., Gierli{\'n}ski M., 2010, MNRAS, 402, 767 

\bibitem[\protect\citeauthoryear{Zdziarski et al.}{2012}]{z12} 
Zdziarski A.~A., Sikora M., Dubus G., Yuan F., Cerutti B., Ogorza{\l}ek A., 2012, MNRAS, 421, 2956

\bibitem[\protect\citeauthoryear{Zdziarski, Miko{\l}ajewska \& Belczy{\'n}ski}{Zdziarski et al.}{2013}]{zmb13}
Zdziarski A.~A., Miko{\l}ajewska J., Belczy{\'n}ski K., 2013, MNRAS, 429, L104

\bibitem[\protect\citeauthoryear{Zdziarski, Segreto \& Pooley}{Zdziarski et al.}{2016}]{z16} Zdziarski A.~A., Segreto A., Pooley G.~G., 2016, MNRAS, 456, 775

\bibitem[\protect\citeauthoryear{Zdziarski et al.}{2018}]{z18} Zdziarski A.~A., et al., 2018, MNRAS, 479, 4399

\bibitem[\protect\citeauthoryear{Zhang et al.}{1997}]{1997ApJ...477L..95Z} Zhang S.~N., Cui W., Harmon B.~A., Paciesas W.~S., Remillard R.~E., van Paradijs J., 1997, ApJ, 477, L95


\end{thebibliography}
\end{document}